\documentclass[floatfix, preprint, onecolumn, prd,showpacs, showkeys, preprintnumbers, nofootinbib, superscriptaddress]{revtex4-1}
\usepackage{times}
\usepackage{amssymb,amsbsy,amsmath,amsfonts}
\usepackage{graphicx}
\usepackage{float}
\usepackage{color}
\usepackage{morefloats}
\usepackage{rotating}
\usepackage{srcltx}
\usepackage{slashed}
\usepackage{latexsym}
\usepackage{subfigure}
\usepackage{multirow}
\usepackage{verbatim}
\usepackage{hyperref}
\usepackage{tabularx}
\usepackage{longtable}
\renewcommand{\arraystretch}{1.2}

\begin{document}

\title{ Octet baryon magnetic moments at next-to-next-to-leading order in covariant chiral perturbation theory}

\author{Yang Xiao}
\affiliation{School of Physics and Nuclear Energy Engineering, Beihang University, Beijing 100191, China}

\author{Xiu-Lei Ren}
\affiliation{State Key Laboratory of Nuclear Physics and Technology, School of Physics, Peking University, Beijing 100871, China}
\affiliation{Institut f\"{u}r Theoretische Physik II, Ruhr-Universit\"{a}t Bochum, D-44780 Bochum, Germany}

\author{Jun-Xu Lu}
\affiliation{School of Physics and Nuclear Energy Engineering, Beihang University, Beijing 100191, China}

\author{Li-Sheng Geng}
\email[E-mail: ]{lisheng.geng@buaa.edu.cn}
\affiliation{School of Physics and
Nuclear Energy Engineering \&
Beijing Key Laboratory of Advanced Nuclear Materials and Physics,  Beihang University, Beijing 100191, China\&
Beijing Advanced Innovation Center for Big Date-based Precision Medicine, Beihang University, Beijing100191, China
}

\author{Ulf-G. Mei\ss{}ner}
\affiliation{Helmholtz-Institut f\"ur Strahlen- und Kernphysik and Bethe Center
for Theoretical Physics, Universit\"at Bonn, D-53115 Bonn, Germany}
\affiliation{Institute for Advanced Simulation and J\"ulich Center for Hadron Physics,
Institut f\"ur Kernphysik, Forschungszentrum J\"ulich, D-52425 J\"ulich, Germany}

\begin{abstract}

We calculate the octet baryon magnetic moments in covariant baryon chiral perturbation theory  with the extended-on-mass-shell 
renormalization scheme up to next-to-next-to-leading order.
At this order, there are nine low-energy constants, which cannot be uniquely determined by the seven experimental data alone. 
We propose two strategies to circumvent this problem.
First, we assume that chiral perturbation theory  has a certain convergence rate and use this as one additional constraint to fix the
low-energy constants  by fitting to the experimental data. Second, we fit to lattice QCD simulations to determine the low-energy constants.
We then compare the resulting predictions of the light and strange quark mass dependence of the octet baryon magnetic moments by the three mostly studied formulations of baryon chiral perturbation theory,
namely, the extended-on-mass-shell, the infrared, and the heavy baryon approach.  It is shown that once more precise lattice data become available, one 
will learn more about the convergence pattern of baryon chiral perturbation theory.

\end{abstract}
\pacs{12.39.Fe, 14.20.Dh, 14.20.Jn,13.40.Em}


\date{\today}

\maketitle
\section{Introduction:}
SU(3) flavor symmetry and its breaking play an important role in our understanding of the strong interaction in the non-perturbative regime.
In the limit of an exact SU(3) flavour symmetry,  one can relate the magnetic moments of the octet baryons and the $\Lambda\Sigma^0$ transition to those of the proton and the neutron via the celebrated Coleman-Glashow formulae~\cite{Coleman:1961jn}.  Nonetheless, in nature SU(3) flavor symmetry is broken. This must be properly taken into account in order to improve the description of the baryon magnetic moments by inducing a realistic SU(3)-breaking mechanism. Chiral perturbation theory (ChPT), the low-energy effective field theory of
QCD (see e.g.~\cite{Gasser:1984gg,Gasser:1987rb,Bernard:1995dp,Scherer:2002tk,Bernard:2007zu}),  provides an appropriate framework to tackle this problem in a systematic fashion.  However, it was  noticed long ago that the leading order (LO) chiral corrections are large and tend to worsen the results, as exemplified, e.g., in~\cite{Caldi:1974ta,Bijnens:1985kj,Jenkins:1992pi,Kubis:2000aa}. This issue has often been used to question the validity of the SU(3) baryon ChPT altogether, see e.g.~\cite{Donoghue:2004vk}.

 It was shown in Ref.~\cite{Geng:2008mf} that  one can achieve, however, an order by order improvement in the description of the octet baryon magnetic moments with the extended-on-mass-shell (EOMS) formulation of baryon  ChPT~\cite{Fuchs:2003qc}~\footnote{Note that  the contribution of the virtual decuplet baryons has an negligible effect on the results~\cite{Geng:2009hh}, consistent with the heavy baryon (HB) findings of
 Ref.~\cite{Bernard:1998gv}.}. Although it seems that the puzzle has been solved, a natural question is what happens at the next-to-next-to-leading order (NNLO). Because of the increased number of unknown low-energy constants (LECs), a clear answer to this question has not yet  been provided.

In the last two decades several calculations of the octet baryon magnetic moments in heavy baryon (HB) ChPT  up to NNLO have been performed both with~\cite{Jenkins:1992pi,Durand:1997ya,Puglia:1999th} and without~\cite{Meissner:1997hn} the inclusion of the baryon decuplet. It was shown in Ref.~\cite{Meissner:1997hn} that
at NNLO the convergence of the HBChPT is quite good, contrary to the pattern exhibited at NLO.  One should note that, however, in Ref.~\cite{Meissner:1997hn} the contributions of the
two NNLO LECs $b_6^{D',F'}$ are absorbed into the two LO ones $b_6^{D,F}$. This is legitimate as long as one works
at the physical quark masses, as the primed LECs merely amount to a quark mass dependent shift of the unprimed ones that can not 
be disentangled.

Despite of all these studies, it remains unclear whether
the convergence pattern of the HB~\cite{Jenkins:1992pi}, the infrared (IR)~\cite{Kubis:2000aa}, and the EOMS formulation~\cite{Geng:2008mf} observed in
the description of the octet baryon magnetic moments, one of the cleanest observables,  is particular to NLO and accidental, where no unknown LECs contribute, or it might be a more genuine feature of different formulations.
Given that ChPT plays an indispensable role in our understanding of low energy strong interaction physics, it is of utmost importance to clarify this puzzling situation.

In this work we address this question by performing a study of the octet baryon magnetic moments at NNLO using the EOMS renormalization scheme. Since at this order the number of LECs is larger than that of the data,
we first use convergence as a  criterion to constrain the two LO  LECs, $b^{D,F}_6$, which cannot be distinguished from the
two NNLO LECs $b_6^{D',F'}$ at the physical point. We then predict the light quark mass dependence of the octet baryon magnetic moments and  contrast them with  the
results of the HB and IR formulations, and the state of the art lattice QCD simulations. Second, we fit the LECs to the lattice QCD data and then predict the strange quark mass dependence of the magnetic moments.  It is shown
that depending on how one determines the LECs, the predicted dependence is rather different, which could be investigated in more detail
by future lattice QCD simulations.

This paper is organized as follows. In Section~\ref{sec:theo}, we present the theoretical framework.  Results and discussions are given in 
Section~\ref{sec:res}, followed by a short summary in Section~\ref{sec:summ}.

\section{Theoretical framework}
\label{sec:theo}
The octet baryon magnetic moments are defined via baryon matrix elements of the electromagnetic
current  $J_\mu$ as follows
\begin{equation}
\langle \bar B|J_\mu|B \rangle=\bar {u}(p_f)\left[ \gamma_{\mu} F_{1}^{B}(t)
+\frac{i\sigma_{\mu\nu}q^{\nu}}{2m}F_{2}^{B}(t)\right]u(p_i),
\end{equation}
where $\bar{u}$ and $u$ are Dirac spinors, $m$ is the baryon mass, and $F_1^B$ and $F_2^B$ denote the
Dirac and Pauli form factor, respectively. The four-momentum transfer is defined as $q=p_f-p_i$  and $t=q^2$.
At $t=0$,  $F_{2}^{B}(0)$  is the so-called anomalous magnetic moment, $\kappa_{B}$, and the magnetic moment is 
$\mu_{B}=\kappa_{B}+ q_{B}$, with $q_B$ the charge of the baryon.

In ChPT, one can calculate the baryon magnetic moments order by order, i.e.,
\begin{equation}
\mu_B=\mu_B^{(2)}+\mu_B^{(3)}+\mu_B^{(4)}+\cdots,
\end{equation}
where the numbers in the superscripts are the chiral order, defined as
 $n_\mathrm{ChPT}=4L-2N_M-N_B+\sum_k k V_k$
for a properly renormalized diagram with $L$ loops, $N_M$($N_B$) meson (baryon) propagators, and $V_k$ vertices from the $k^{\rm th}$ order Lagrangian. Because of
the large non-zero baryon mass in the chiral limit, this power counting is broken in a naive
application of the $\overline{\mathrm{MS}}$ regularization scheme~\cite{Gasser:1987rb}.
To recover the power counting, several approaches have been proposed, such as
the HB method~\cite{Jenkins:1990jv,Bernard:1992qa}, the IR approach~\cite{Becher:1999he} and
the EOMS scheme~\cite{Fuchs:2003qc}.
In recent years, it has been shown that the EOMS scheme has some advantages because
it satisfies all symmetry and  analyticity constrains and converges relatively faster in certain cases,
see e.g. Ref.~\cite{Geng:2013xn} for a short review.

\begin{figure}[t!]
  \centering
  \includegraphics[width=0.5\textwidth]{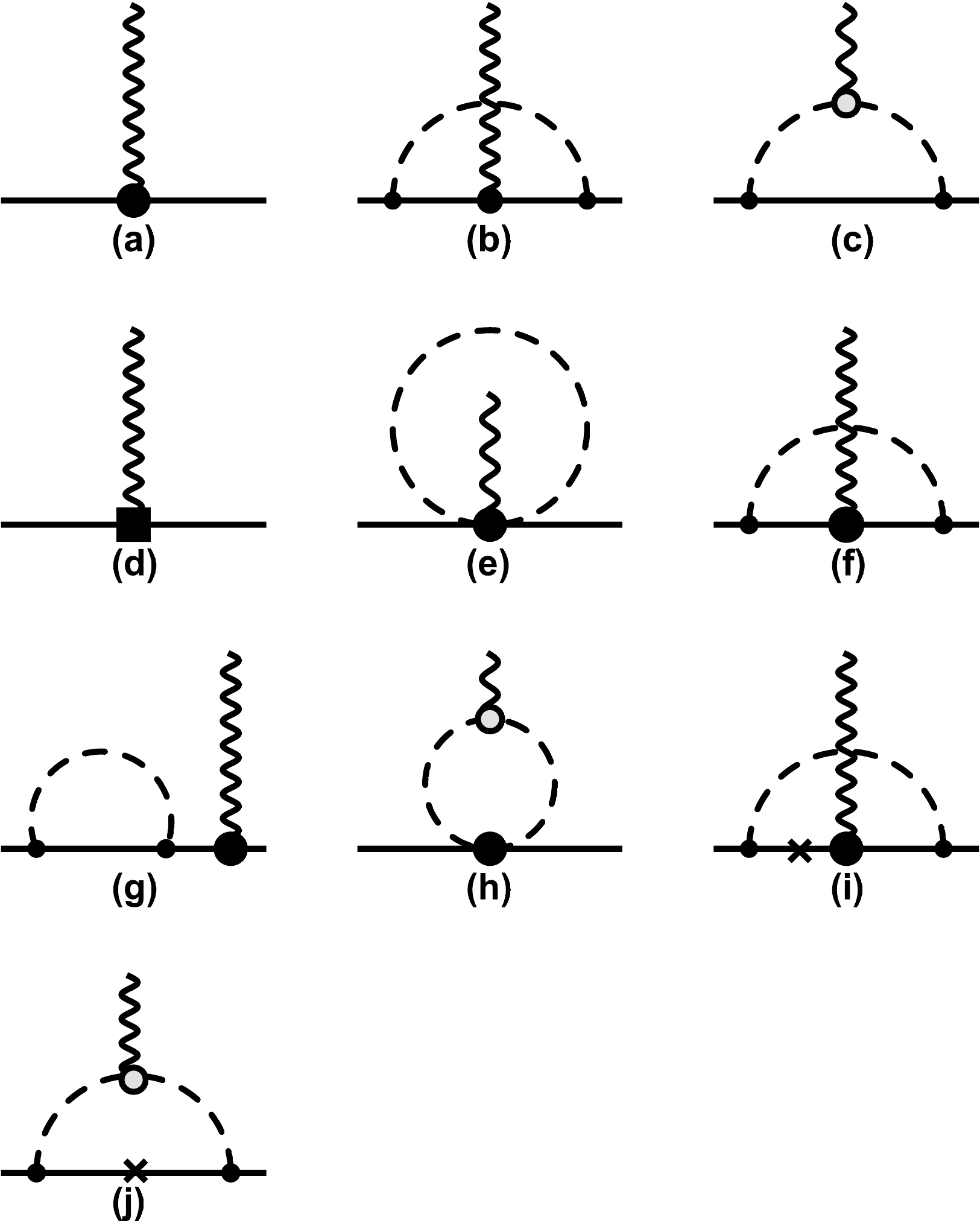}
  \caption{Feynman diagrams contributing to the octet baryon magnetic moments up to NNLO. 
  (a) contributes to LO, while (b) and (c) depict the NLO corrections. All other diagrams represent the
  NNLO contribution.
  Solid, dashed and wiggly lines denote baryons, Goldstone bosons and photons, respectively.
  The small and  medium solid dots refer to vertices obtained from $\mathcal{L}_{MB}^{(1)}$ and
  $\mathcal{L}_B^{(1)}$, in order. The heavy dots refer to vertices from $\mathcal{L}_{MB}^{(2)}$ .
  The circles refer to vertices from $\mathcal{L}_{M}^{(2)}$.
  The crosses refer to mass insertions $\sim$ $b_{D,F}$. The square refers to a vertex
   from  $\mathcal{L}_{MB}^{(4)}$.  Crossed graphs are not shown.}
  \label{fig:feynman}
\end{figure}
The Feynman diagrams needed to calculate $\mu_B$ up to NNLO are shown in Fig.~\ref{fig:feynman}.
The LO contributions are provided by the following Lagrangian,
\begin{equation}
\mathcal{L}_{MB}^{(2)}=\frac{b_{6}^{D}}{8m}\langle \bar{B}\sigma^{\mu\nu}\left\{F_{\mu\nu}^{+},B\right\}\rangle+\frac{b_{6}^{F}}{8m}\langle \bar{B}\sigma^{\mu\nu}\left[F_{\mu\nu}^{+},B\right]\rangle,
\end{equation}
where $\sigma^{\mu\nu}=\frac{i}{2}\left[\gamma^{\mu} ,\gamma^{\nu}\right]$,
$F_{\mu\nu}^{+}=(u^{\dag}QF_{\mu\nu}u+uQF_{\mu\nu}u^{\dag})$ , $Q=|e|\mathrm{diag}(2,-1,-1)/3$
is the quark charge matrix,  $u= \mathrm{exp}\left[i\Phi/2F_\phi\right]$, with $\Phi$ the unimodular
matrix containing the pseudoscalar nonet, $F_\phi$ the pseudoscalar decay constant, and
$F_{\mu\nu}=\partial_{\mu}A_{\nu}-\partial_{\nu}A_{\mu}$ is the conventional photon field
strength tensor.  Moreover, $\langle \ldots \rangle$ denotes the trace in flavor space.
At LO, there are only two LECs, $b_{6}^{D}$ and $b_{6}^{F}$.

The leading SU(3) breaking corrections come from loop diagrams. They  arise at order $\mathcal{O}(p^3)$
in the chiral counting and
are determined completely in terms of the lowest order LECs from $\mathcal{L}^{(1)}_B+\mathcal{L}^{(2)}_{M}
+ \mathcal{L}^{(1)}_{MB}$, namely,
\begin{equation}
\mathcal{L}_{B}^{(1)}=\langle \bar{B}i\gamma^{\mu} D_\mu B - m_0 \bar B B\rangle,
\end{equation}
\begin{equation}
\mathcal{L}_{M}^{(2)}=\frac{F_\phi^{2}}{4}\langle u_{\mu}u^{\mu}+\chi^{+}\rangle,
\end{equation}
\begin{equation}
\mathcal{L}_{MB}^{(1)}=\frac{D}{2}\langle \bar{B}\gamma^{\mu}\gamma^{5}\left\{u_{\mu},B\right\}\rangle+\frac{F}{2}\langle \bar{B}\gamma^{\mu}\gamma^{5}\left[u_{\mu},B\right]\rangle,
\end{equation}
where $m_0$ denotes the baryon mass in the chiral limit, $D_\mu B=\partial_\mu B+[\Gamma_\mu, B]$,  $\Gamma_\mu=\frac{1}{2}(u^\dagger\partial_\mu u
+ u\partial_\mu u^\dagger)-\frac{i}{2}(u^\dagger v_\mu u + u v_\mu u^\dagger)$, with $v_\mu$  the vector
source, $u_{\mu}=i(u^{\dag}\partial_{\mu}u-u\partial_{\mu}u^{\dag})+(u^\dagger v_\mu u- uv_\mu u^
\dagger)$ and $\chi^{\pm}=u^{\dag}\chi u^{\dag}\pm u\chi u$. Here, $\chi=2B_{0}\mathcal{M}$,
with $\mathcal{M}$ the quark mass matrix $\mathcal{M}=\mathrm{diag}(m_q,m_q,m_s)$. In what follows,
we work in the isospin limit $m_q=(m_u+m_d)/2$. Further,
$B_{0}=|\langle0|\bar{q}q|0\rangle|/F_\phi^{2}$. The axial vector couplings $D$ and $F$ are
determined from hyperon decays to be $D=0.8$ and $F=0.46$, and these values will be taken
throughout.

The explicit expressions of the LO and NLO results can be found in Ref.~\cite{Geng:2008mf}.
In the following, we focus on the NNLO contributions. At this order one has to include
one-loop diagrams with exactly one vertex from $\mathcal{L}^{(2)}_{MB}$ as well as additional tree
contributions from $\mathcal{L}^{(4)}_{MB}$. The fourth order contribution to $\mu_B$ is
given as  (see Fig.~1):
\begin{equation}
\mu^{(4)}=\mu^{(4,d)}+\mu^{(4,e)}+\mu^{(4,f)}+\mu^{(4,g)}+\mu^{(4,h)}+\mu^{(4,i)}+\mu^{(4,j)}.
\end{equation}

The terms contributing to $\mu^{(4,d)}_{B}$ collect the tree contributions with exactly
one insertion from the following chiral Lagrangian~\cite{Meissner:1997hn,Jiang:2016vax},
\begin{equation}
 \begin{split}
  \mathcal{L}_{MB}^{(4)}=+&\frac{b_6^{D'}}{8m}\langle\chi^{+}\rangle \langle \bar{B}\sigma^{\mu\nu}\{F_{\mu\nu}^{+},B\}\rangle+\frac{b_6^{F'}}{8m}\langle\chi^{+}\rangle \langle \bar{B}\sigma^{\mu\nu}[F_{\mu\nu}^{+},B]\rangle\\
+&\frac{\alpha_1}{8m}\langle \bar{B}\sigma^{\mu\nu}\left[\left[F^{+}_{\mu\nu},B\right],\chi^{+}\right]\rangle+\frac{\alpha_2}{8m}\langle \bar{B}\sigma^{\mu\nu}\left\{\left[F^{+}_{\mu\nu},B\right],\chi^{+}\right\}\rangle\\
  +&\frac{\alpha_3}{8m}\langle\bar{B}\sigma^{\mu\nu}\left[\left\{F^{+}_{\mu\nu},B\right\},\chi^{+}\right]\rangle
  +\frac{\alpha_4}{8m}\langle\bar{B}\sigma^{\mu\nu}\left\{\left\{F^{+}_{\mu\nu},B\right\},\chi^{+}\right\}\rangle\\
  +&\frac{\beta_1}{8m}\langle \bar{B}\sigma^{\mu\nu}B\rangle\langle\chi^{+}F_{\mu\nu}^{+}\rangle, \end{split}
\end{equation}
where $\alpha_{1,2,3,4}$ and $\beta_1$ are LECs.

At this order, we also have to consider double derivative operators at the meson-baryon vertex
with the photon hooking on to the
meson loop, see Fig.~1(h). The corresponding terms of the dimension two Lagrangian
read~\cite{Meissner:1997hn,Jiang:2016vax}
\begin{equation}
\mathcal{L}_{MB}^{(2')}=\frac{i}{2}\left\{b_{9}\langle \bar {B}\sigma^{\mu\nu}u_\mu\rangle\langle
u_\nu B\rangle+b_{10,11}\langle \bar {B}\sigma^{\mu\nu}\left(\left[u_\mu,u_\nu\right],B\right)_{\pm}\right\rangle\},
\end{equation}
where $b_{9,10,11}$ are LECs.  They are estimated via resonance saturation in Ref.~\cite{Meissner:1997hn} and
re-evaluated at $m_0=0.94$ GeV, yielding $b_9=0.43$ GeV$^{-1}$, $b_{10}=0.86$ GeV$^{-1}$, and $b_{11}=0.45$ GeV$^{-1}$. We will call these values of the LECs set~I. An improved determination has been
given in Ref.~\cite{Kubis:2000aa}, the corresponding values are  $b_9=1.36$ GeV$^{-1}$, 
$b_{10}=1.24$ GeV$^{-1}$, and $b_{11}=0.46$ GeV$^{-1}$. We will refer to these as set~II.

At NNLO, one also needs the LO chiral corrections to the baryon masses, which are provided by
the following Lagrangian:
\begin{equation}
\mathcal{L}_{MB}^{(2'')}=b_D\langle \bar {B}\left\{\chi_{+},B\right\}\rangle+b_F\langle \bar {B}\left[\chi_{+},B\right]\rangle.
\end{equation}
The two LECs $b_D$ and $b_F$ are fixed from the octet baryon mass splittings, yielding $b_D=0.066$~GeV$^{-1}$ and $b_F=-0.21$~GeV$^{-1}$.

The NNLO tree-level contributions can be obtained rather straightforwardly. The results are
shown in the Appendix.
Following the EOMS prescription to restore the power counting, we obtain the following NNLO loop results,

\begin{equation}
\mu_B^{(4,e)}(m_\phi)=C_B^{(4,e)}(\phi)\frac{m_\phi^2}{16\pi^2 F_{\phi}^2}\log\left(\frac{m_\phi^2}{\mu^2}\right),~~~~~~~~~~~~~~~~~~~~~~~~~~~~~~~~~~~~~~~~ ~~~~~~~~~~~~~~~~~~~~~~~~~~~~~~~~~~~~~~~~~~~~~~~~~~~~~~~~~~~~~~~~~~~~~~~~
\end{equation}
\begin{equation}
\begin{split}
\mu_{B}^{(4,f)}(m_\phi)=&\frac{-C_{B}^{(4,f)}(\phi)}{16 \pi ^2 F_{\phi }^2 m^2 \sqrt{4 m^2-m_\phi^2}}\left\{\left(2 m_\phi^5-4 m^2 m_\phi^3\right) \cos^{-1} \left(\frac{m_\phi}{2m}\right)\right.~~~~~~~~~~~~~~~~~~~~~~~~~~~~~~~~~~~~~~~~\\
   &\left.+\sqrt{4 m^2-m_\phi ^2} \left\{m^2m_\phi ^2 \left[\log
   \left(\frac{m_\phi ^2}{\mu ^2}\right)+2\right]-m_\phi ^4\log \left(\frac{m_\phi^2}{m^2}\right)\right\}\right\},
\end{split}
\end{equation}
\begin{equation}
\begin{split}
 \mu_{B}^{(4,g)}(m_\phi)=
 &\frac{C_{B}^{(4,g)}(\phi) m_\phi^2}{16\pi ^2 F_\phi ^2 m^2 \left(4 m^2-m_\phi^2\right)} \left\{\left(4 m^2-m_\phi^2\right) \left\{m^2 \left[\log \left(\frac{\mu ^2}{m^2}\right)-4\right]\right.\right.~~~~~~~~~~~~~~~~~~~~~~~~~~~~~~~~~~~~~~~~~~~~~~~~~~~~~~~~~~~~~~~\\
 &\left.\left.+\left(2 m_\phi^2-3 m^2\right) \log \left(\frac{m_\phi^2}{m^2}\right)\right\}+4 m_\phi\left(3 m^2-{m_\phi}^2\right) \sqrt{4 m^2-m_\phi^2} \cos ^{-1}\left(\frac{m_\phi}{2m}\right)\right\},
\end{split}
\end{equation}
\begin{equation}
\mu_{B}^{(4,h)}(m_\phi)=C_{B}^{(4,h)}(\phi)\frac{mm_{\phi}^2 \log\left(\frac{m_{\phi}^2}{\mu ^2}\right)}{8 \pi ^2 F_{\phi}^2},~~~~~~~~~~~~~~~~~~~~~~~~~~~~~~~~~~~~~~~~~~~~~~~~~~~~~~~~~~~~~~~~~~~~~~~~~~~~~~~~~~~~~~~~~~~~~~~~~~~~~~~~~~~~~~
\end{equation}
\begin{equation}
\begin{split}
\mu_{B}^{(4,i)}(m_\phi)=&\frac{C_{B}^{(4,i)}(\phi)}{4 \pi ^2 m^3F_{\phi }^2 \left(4 m^2-m_\phi ^2\right)^{3/2}}\left\{ 2m_\phi\left(-16 m^6+30 m^4m_\phi ^2-10 m^2 m_\phi ^4+m_\phi^6\right)\cos ^{-1}\left(\frac{m_\phi}{2m}\right)\right.~~~~~~~~~~~~~~~~~~~~~~~~~~~~\\
                &\left.+\sqrt{4 m^2-m_\phi ^2} \left[4m^6-13 m^4 m_\phi ^2+2 m^2 m_\phi^4-\left(m_\phi ^3-4 m^2 m_\phi\right)^2 \log \left(\frac{m_\phi^2}{m^2}\right)\right]\right\},
\end{split}
\end{equation}
\begin{equation}
\begin{split}
\mu_{B}^{(4,j)}(m_\phi)=&\frac{C_{B}^{(4,j)}(\phi)}{4\pi^2F_\phi^2m^3\left(4m^2-m_\phi^2\right)^{3/2}}\left\{ \sqrt{4m^2-m_\phi^2}\left[-20m^6+15m^4m_\phi ^2-2 m^2 m_\phi ^4\right.\right.~~~~~~~~~~~~~~~~~~~~~~~~~~~~~~~~~\\
                &\left.\left.-\left(4 m^6-21m^4 m_\phi ^2+9 m^2 m_\phi^4-m_\phi ^6\right) \log\left(\frac{m_\phi^2}{m^2}\right) \right]\right.\\
                &\left.-2m_\phi \left(-32 m^6+37 m^4m_\phi ^2-11 m^2 m_\phi ^4+m_\phi ^6\right) \cos ^{-1}\left(\frac{m_\phi}{2m}\right)\right\},
\end{split}
\end{equation}
where the coefficients $C_B^{(4,e,f,h,i,j)}(\phi=\pi, K, \eta)$  are tabulated in the Appendix.  We have checked that our results agree with those of Ref.~\cite{Meissner:1997hn} in the heavy mass limit up to analytical terms.

\section{Results and discussions}
\label{sec:res}
At LO and NLO, the two LECs $b_6^{D,F}$ can be determined by fitting to the seven experimental
data $\mu_{p,n,\Lambda, \Sigma^+,\Sigma^-,
\Xi^{-},\Xi^{0}}$. An extensive discussion of the EOMS results in comparison with
the HB and IR results is given in Ref.~\cite{Geng:2008mf}. At NNLO, however, there are nine LECs,
two from the LO contribution and seven from the NNLO contribution. As a result,
the experimental data alone can not uniquely determine all the nine LECs. In Ref.~\cite{Meissner:1997hn},
the two NNLO LECs $b_{6}^{D',F'}$ are absorbed by the two LO LECs,
$b_{6}^{D,F}$, while in the present work we keep explicitly  $b_6^{D',F'}$.
Note that at the physical point, only the combinations $\bar{b}^{D,F}_6=b_{6}^{D,F} + \langle \chi_+\rangle b_6^{D',F'}$
are relevant.
In our numerical analysis,
the decay constant and the chiral limit value of the baryon masses are chosen to be $F_{\phi}=0.108$~GeV,
an average of the pion, kaon, and eta decay constants,  and $m_0=0.94$~GeV, following the argument in
 Ref.~\cite{Meissner:1997hn}. 
The  dimensional regularization scale is set at $\mu=1.0$ GeV. We also have
performed calculations allowing for  slight variation of $m$ or $\mu$ about these values,  e.g., $\mu=0.9\sim1.1$ GeV and $m=0.8\sim1.1$ GeV, and found that such changes  have negligible effects on
our results. 

\begin{figure}[htpb]
  \centering
\centering
\includegraphics[height=3in,width=3in]{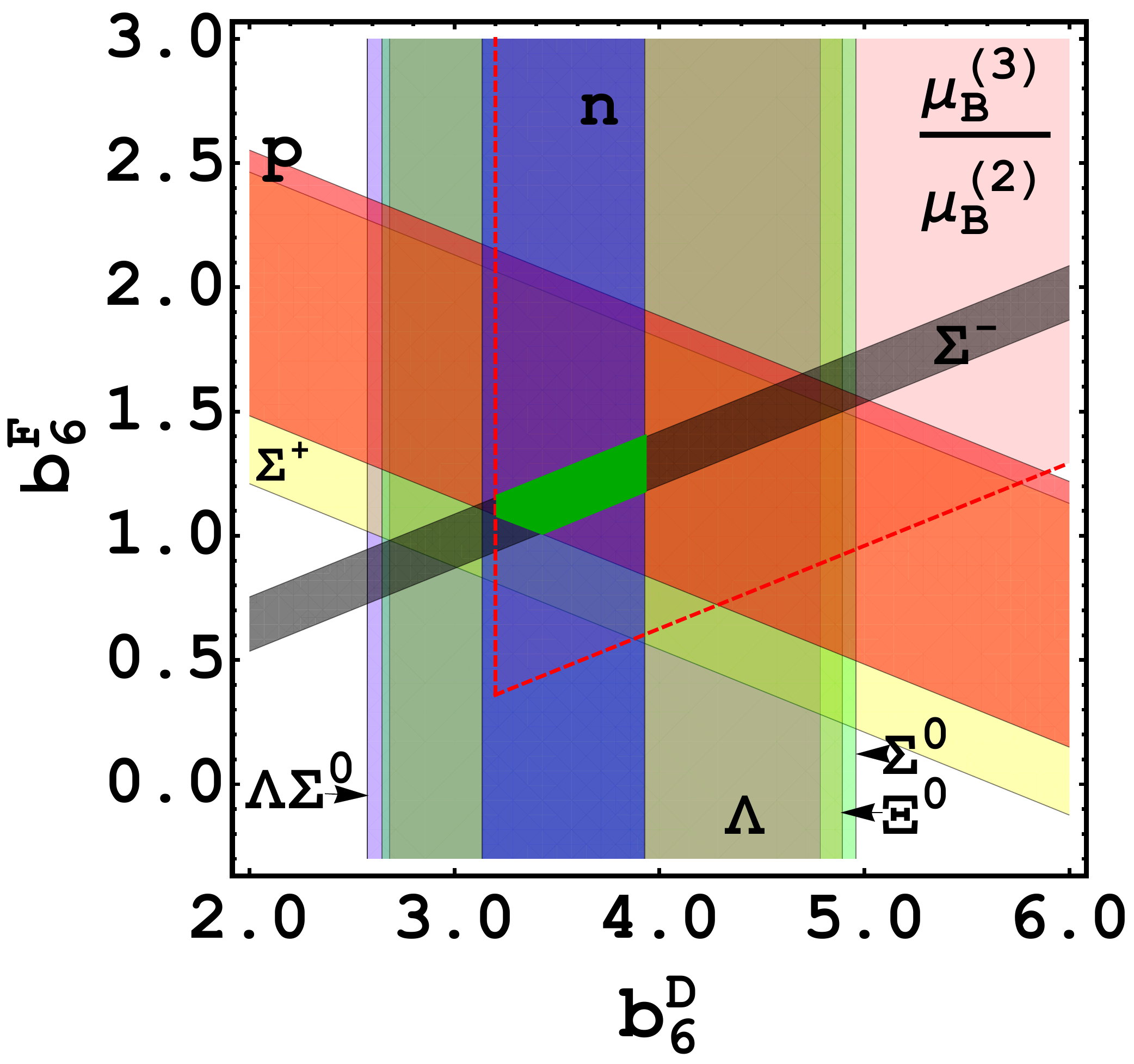}\;\;\;\;
\includegraphics[height=3.in,width=3.in]{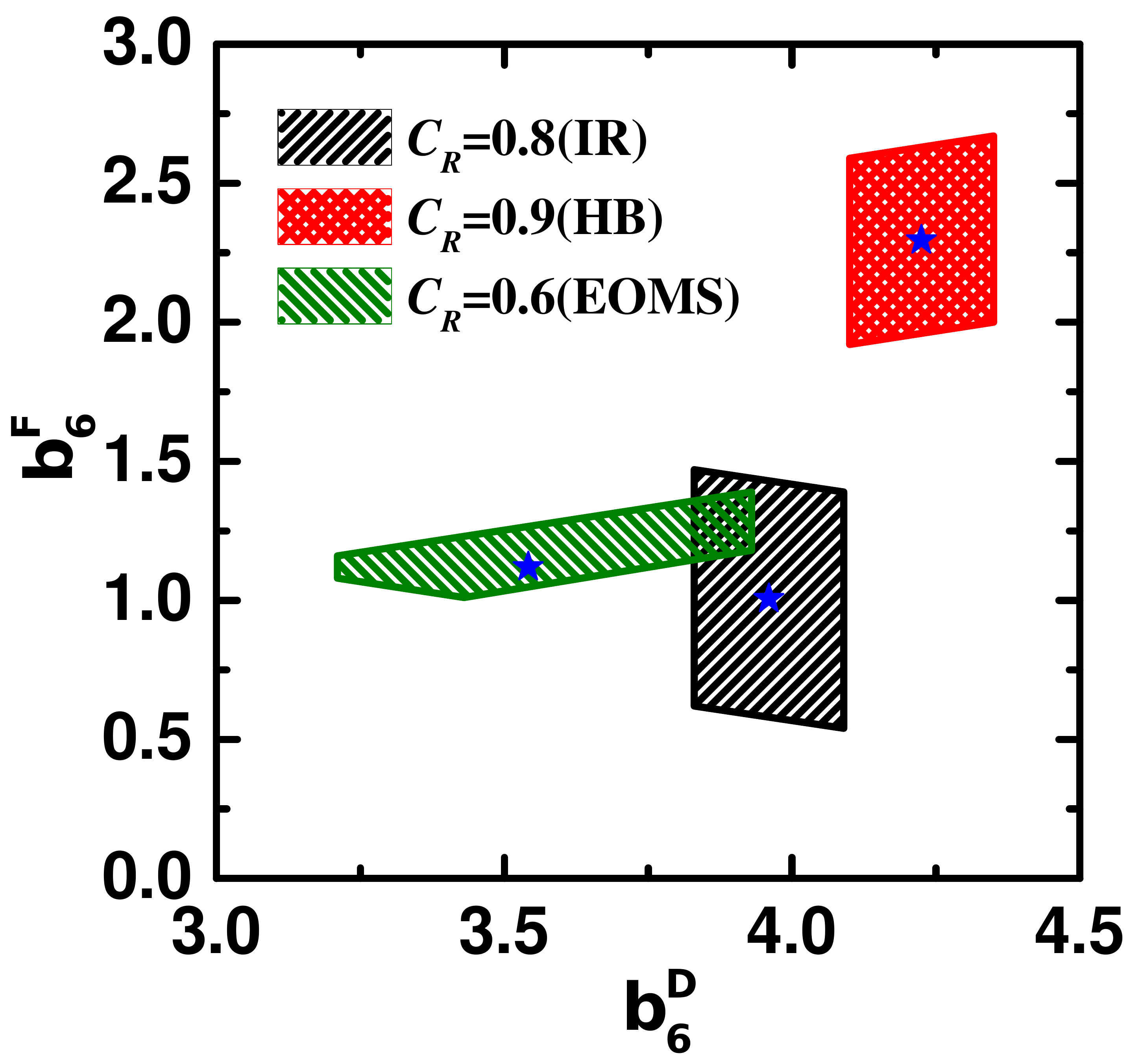}
  \caption{Left panel:  dependence of the convergence rate $C_R=0.6$  on  $b_6^{D,F}$ in the EOMS scheme.
  Right panel: convergence rates, $C_R$, of the HB, IR, and EOMS scheme as a function of $b_6^{D,F}$. The green shaded area in both plots are the same.
  The blue stars denote the optimal $b_6^{D,F}$, located in the center of the shaded area and later used to study convergence,   i.e., $(b_6^{D}, b_6^{F})=(3.54,1.17)_\mathrm{EOMS}$, $(4.22, 2.30)_\mathrm{HB}$, $(3.96, 1.00)_\mathrm{IR}$.}
  \label{fig:conv1}
\end{figure}

\subsection{Using convergence to constrain the low-energy constants}
We assume that BChPT has a reasonable convergence rate in the $u,d,s$ three flavor
sector, namely, higher order contributions are suppressed compared to lower order ones,
in terms of $M_K/\Lambda_\mathrm{ChPT}\approx0.5$. Under this assumption, we can use the convergence criterion and
the experimental data to constrain the nine LECs. More specifically, we define the convergence
rate ($C_R$) as
\begin{equation}
C_R = \mathrm{max}(\mu_B^{(3)}/\mu_B^{(2)},\mu_B^{(4)}/\mu_B^{(3)})\quad\mathrm{with}\quad B=p,n,
\Lambda, \Sigma^+,\Sigma^0,\Sigma^-,\Lambda\Sigma^0,
\Xi^{-},\Xi^{0}.
\end{equation}

\begin{table}[htpb]
\centering
\caption{Contributions of different chiral orders of the HB, IR, and EOMS schemes up to NNLO with the LECs of Table \ref{tab:LEC1}. }
\begin{tabular}{ccccccc}
 \hline
 \hline
 \multirow{2}{*}{Baryons} &\multicolumn{2}{c}{EOMS}            &\multicolumn{2}{c}{IR}          &\multicolumn{2}{c}{HB}\\
 \cline{2-7}
                          &$\mu_B^{(3)}/\mu_B^{(2)}$        &$\mu_B^{(4)}/\mu_B^{(3)}$             &$\mu_B^{(3)}/\mu_B^{(2)}$         &$\mu_B^{(4)}/\mu_B^{(3)}$           &$\mu_B^{(3)}/\mu_B^{(2)}$          &$\mu_B^{(4)}/\mu_B^{(3)}$  \\
\hline
  $p$                    &$-0.27$          &$-0.38$               &$-0.16$         &$0.01$            &$-0.44$	        &$-0.07$\\
  $n$                      &$-0.19$	      &$0.02$	           &$-0.17$	      &$0.61$            &$-0.18$	        &$0.74$\\
 $\Lambda$                &$-0.52$	      &$-0.08$	           &$-0.73$	      &$-0.27$	       &$-0.83$	        &$-0.32$\\
 $\Sigma^{-}$             &$0.18$	          &$-0.04$              &$2.58$	      &$-0.73$	       &$-0.30$	        &$0.30$\\
 $\Sigma^{+}$             &$-0.31$          &$-0.15$	           &$-0.05$	      &$4.20$          &$-0.61$	        &$-0.22$\\
 $\Sigma^{0}$             &$-0.52$	      &$-0.13$               &$-0.73$	      &$-0.31$       &$-0.83$	        &$-0.35$\\
 $\Xi^{-}$                &$0.03$   	      &$-12.88$	           &$3.10$	      &$-1.02$	       &$-0.74$	        &$-0.12$\\
 $\Xi^{0}$                &$-0.54$	      &$-0.13$               &$-0.77$	      &$-0.32$       &$-0.87$	        &$-0.36$\\
 $\Lambda \Sigma^{0}$     &$-0.31$	      &$0.27$	               &$-0.38$	      &$-0.11$       &$-0.43$	        &$0.46$\\
\hline
\hline
\end{tabular}\label{tab:conv1}
\end{table}

When fitting, we can set the convergence rate to a particular value, e.g., 0.5, and search for a
combination of LECs that can satisfy such a requirement.  In the left panel
of Fig.~\ref{fig:conv1}, we show how one can fix the range of $b_6^{D,F}$ by
requiring $C_R\le0.6$ in the EOMS scheme. It can be seen that
 indeed there exist some combinations of $b_6^{D,F}$ which can satisfy the
requirement, namely the green shaded area.
 In the right panel of Fig.~\ref{fig:conv1}, we show the best convergence rate
achievable in the EOMS scheme, in comparison with those in
 the HB and IR schemes. One can see that only in the EOMS scheme a convergence rate about
0.6 can be achieved,~\footnote{ Note that in the searches for the best convergence rate, in the
EOMS scheme  the $\Xi^-$ channel is excluded because of its accidentally tiny contribution  at
NLO. In the same manner, in the IR scheme the $\Sigma^\pm$ and $\Xi^-$ channels are also excluded .}
while the convergence rates in the HB and IR schemes are relatively larger.
 This is consistent
with the findings at NLO~\cite{Geng:2008mf}.
  On the other hand,  from a typical convergence pattern of the three renormalization schemes
shown in Table \ref{tab:conv1} with the corresponding LECs given in Table \ref{tab:LEC1}, one
notices that the contributions of different chiral orders in the HB ChPT are more moderate,
though not as small as one naively expects.

It should be noted that in the numerical study we have taken the values given above for the LECs $D$, $F$, $b_D$, $b_F$,  $b_9$, $b_{10}$, and $b_{11}$. In principle, there are some uncertainties. These can originate either from the data used to fix them, or from the chiral orders at which they are determined,  or  from the validity of the assumption adopted, such as  resonance saturation for the case of $b_9$, $b_{10}$, and $b_{11}$. Unfortunately,
there is no easy way to quantify these uncertainties. We have checked that using set~II instead
of  set~I, we find some quantitative differences but the overall trends  are not affected.

\begin{table}[htpb]
\caption{LO and NNLO low-energy constants denoted by the blue stars in Fig.~\ref{fig:conv1},
where $\bar{b}^D_6$ and $\bar{b}^F_6$ are linear combinations of the two LO and NNLO LECs,
$\bar{b}^D_6=b_6^{D}+\langle \chi_+\rangle b_6^{D'}$, $\bar{b}^F_6=b_6^{F}+\langle \chi_+\rangle b_6^{F'}$.}
\begin{tabular}{cccccccccccc}
  \hline
  \hline
  Chiral Schemes     &$b_6^D$   &$b_6^F$   &$\alpha_1$    &$\alpha_2$   &$\alpha_3$   &$\alpha_4$   &$\beta_1$   &$b_6^{D'}$    &$b_6^{F'}$  &  $\bar{b}^D_6$ & $\bar{b}^F_6$\\
  \hline
   IR                &3.96     &1.00     &$-$0.89    &$-$0.57    &$-$0.19   &0.41   &$-$2.98     &$-$3.85     &$-$1.21 &0.04 &$-$0.23\\
   HB                &4.22     &2.30     &$-$1.49    &0.07     &$-$1.27   &1.74     &$-$2.41     &$-$2.79     &$-$0.82    &1.38   &1.46\\
   EOMS              &3.54     &1.17     &0.08     &0.48     &0.08    &0.83     &$-$0.90     &$-$1.69     &-0.21   &1.82   &0.95\\
  \hline
  \hline
\end{tabular}\label{tab:LEC1}
\end{table}

\begin{figure}[htpb]
  \centering
  \includegraphics[width=1.0\textwidth]{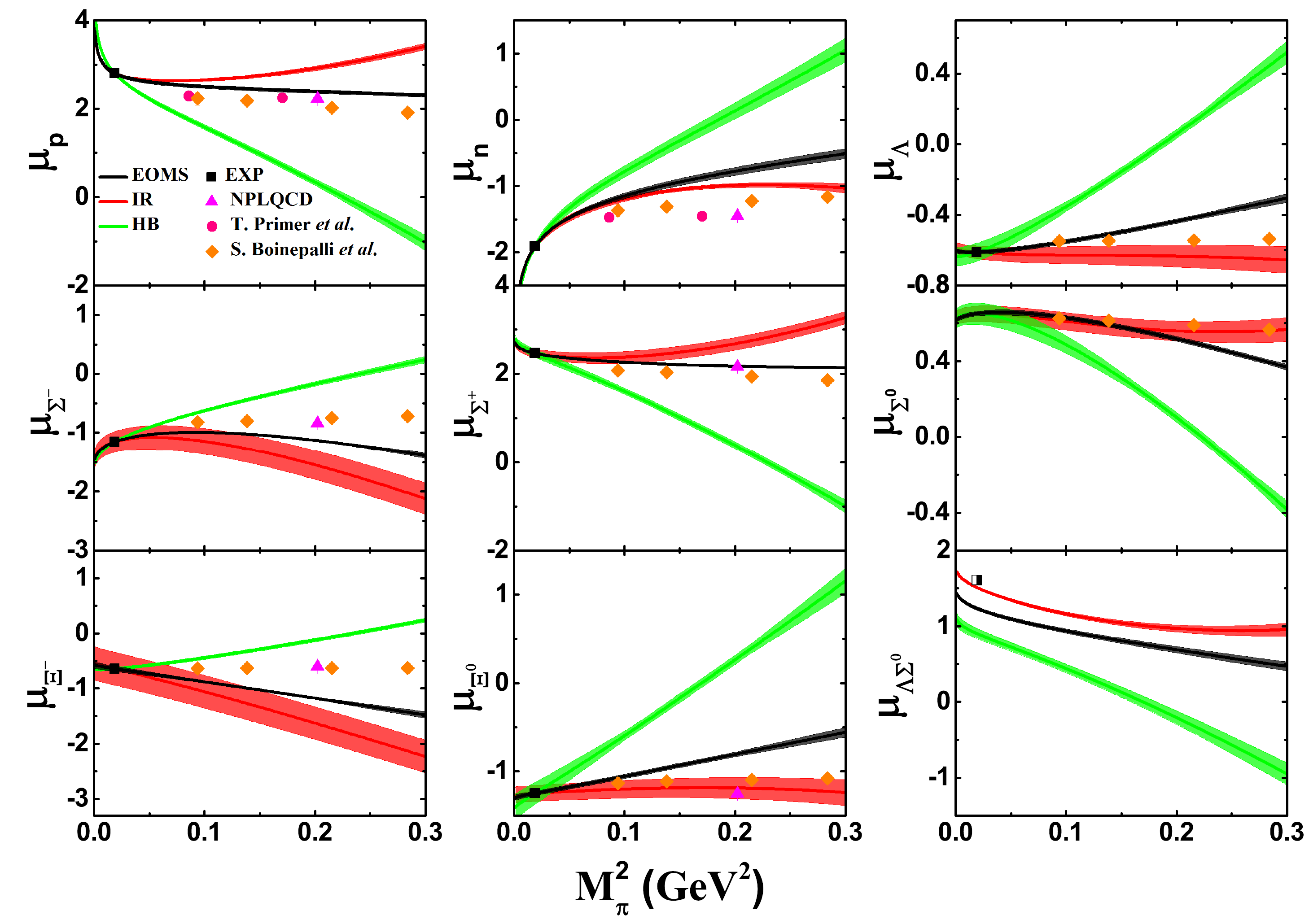}
  \caption{Octet baryon magnetic moments as functions of  the pion mass squared. The central
  values of the LECs are
   given in Table II, while the shaded bands represent higher order contributions not considered (see text for details). The half filled square indicates that
   we have chosen a positive sign for $\mu_{\Lambda\Sigma^0}$.}
  \label{fig:pionmd}
\end{figure}

One way to test the LECs determined above and also to distinguish different formulations
of BChPT is to study the light quark mass dependence of
the magnetic moments. Fixing the strange quark mass to its physical value with the LO
ChPT relation $2 B_0 m_s=(2 M_K^2-M_\pi^2)|_\mathrm{phys.}$, we show the pion mass dependence
of the magnetic moments in Fig.~\ref{fig:pionmd}.  The lines are obtained with the LECs tabulated in Table II, while the bands denote higher order contributions not considered in
the NNLO study. They are obtained according to Ref.~\cite{Binder:2015mbz}:
\begin{equation}
\delta \mu_{B}^{(i)}={\rm {max}} \left( Q^{i-1}|\mu_B^{(2)}|, Q^{i+1-j}|\mu_B^{(j)}| \right),2\leq j\leq i ,
\end{equation}
where $\delta \mu_B^{(i)}$ are the uncertainties of chiral order $i$, $Q=m_\pi/\Lambda_{\rm{QCD}}$,  $m_\pi=0.138~{\rm{GeV}}$, and $\Lambda_{\rm{QCD}}=1$ GeV. In addition, the requirement~\cite{Binder:2015mbz}
\begin{equation}
\delta \mu_B^{(i+1)}\geq Q\delta \mu_B^{(i)}
\end{equation}
is also satisfied. It is clear that these contributions reflect partly the convergence pattern, i.e., the slower the convergence, the larger the higher order contributions.

One can see that the three schemes
display rather different pion mass dependence. 
For comparison,  the state of the art lattice QCD results~\cite{ Boinepalli:2006xd,Primer:2013pva, Parreno:2016fwu} are shown as well. Note that such a comparison is only meant to be qualitative, since these simulations are performed with the strange quark mass close to its physical value but not exactly at the physical point and furthermore current lattice QCD simulations still contain systematic uncertainties not under complete control. It is clear that 
the EOMS results are in better agreement with the lattice data, as corroborated by the unweighted $\tilde{\chi}^2=\sum\left(\mu_\text{th}-\mu_\text{lattice}\right)^2$ between the results of each scheme and the lattice QCD data~\cite{ Boinepalli:2006xd,Primer:2013pva, Parreno:2016fwu} shown in Table~\ref{tab:chi2}. One should note that only at relatively large pion masses, e.g., $M_\pi>200$ or 300~MeV, one can distinguish the results from different formulations of BChPT using lattice QCD simulations.

\begin{table}[htpb]
\centering
\caption{$\tilde{\chi}^2$ between the ChPT results obtained in the HB, IR, and EOMS  schemes and  the LQCD data of Refs.~\cite{ Boinepalli:2006xd,Primer:2013pva, Parreno:2016fwu}with the optimal $b_6^{(D,F)}$ denoted
by the blue stars in Fig.~\ref{fig:conv1}.
}\label{}
\begin{tabular}{cccc}
  \hline
  \hline
  \multirow{2}{*}{Chiral schemes}     &\multicolumn{3}{c}{$\tilde{\chi}^2$}\\
  \cline{2-4}
    &S. Boinepalli et al.~\cite{ Boinepalli:2006xd}     &NPLQCD~\cite{Parreno:2016fwu}     &T. Primer et al.~\cite{Primer:2013pva} \\
  \hline
         IR    &12.98    &2.58     &0.69\\
  HB    &43.45    &12.5     &4.64\\
   EOMS  &3.66     &1.11     &0.50\\

  \hline
  \hline
\end{tabular}\label{tab:chi2}
\end{table}

\subsection{Fitting to the lattice QCD data}
 \begin{figure}[t]
  \centering
  \includegraphics[width=0.8\textwidth]{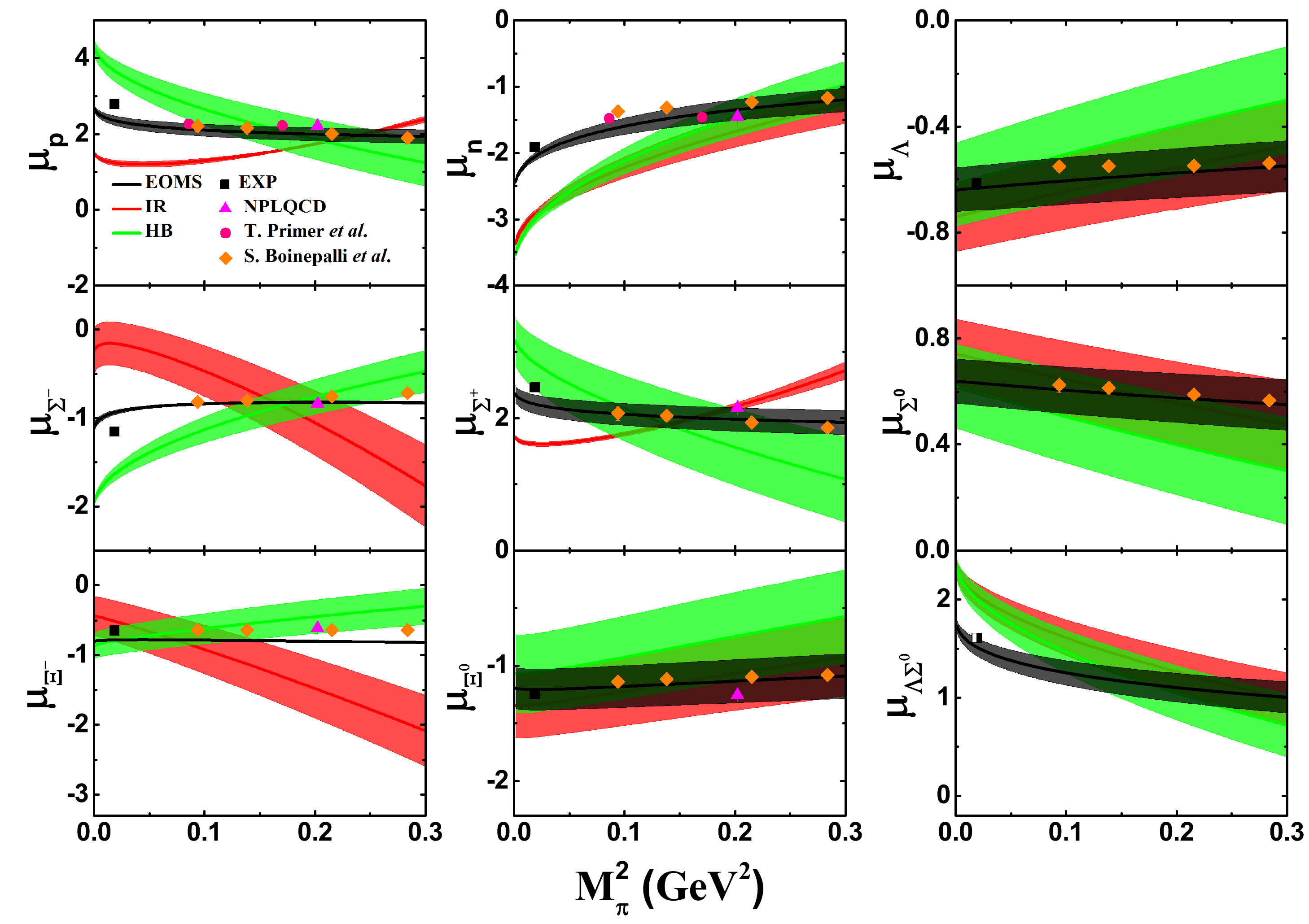}
  \caption{Fits to the lattice QCD data of Ref.~\cite{ Boinepalli:2006xd} at $\mathcal{O}(p^3)$ with the EOMS, HB, and IR BChPT.
  The bands represent estimated higher order corrections as explained in the text.}
  \label{fig:fitlqcdop3}
\end{figure}
\begin{table}[b]
\renewcommand{\tabcolsep}{0.4cm}
\caption{Low-energy constants determined by  fitting to the lattice QCD data of Ref.~\cite{Boinepalli:2006xd} and the corresponding $\tilde{\chi}^2$ in
the IR, HB, and EOMS  BChPT up to NLO.}
\begin{tabular}{ccccccccccc}
  \hline
  \hline
    &$b_6^D$   &$b_6^F$     &$\tilde{\chi}^2$\\
  \hline
   IR                &5.07     &$-$0.92    &13.18   \\
   HB                &5.29     &2.95     &5.75   \\
   EOMS              &3.73     &1.00     &0.61    \\
  \hline
  \hline
\end{tabular}\label{tab:op3fit}
\end{table}

Now we take a more practical attitude, forgetting about the convergence constraint,
and determine the relevant LECs by fitting to the lattice QCD data directly.
Among the three lattice QCD studies we considered, the one of Ref.~~\cite{Boinepalli:2006xd}
features the largest number of simulation points. Therefore, we fit  to these data
to determine the LECs. At $\mathcal{O}(p^3)$,
one has only two LECs.  Their values from the best fit together with the corresponding
$\tilde{\chi}^2$ are tabulated in Table~\ref{tab:op3fit}. The predicted pion mass dependence
is shown in Fig.~\ref{fig:fitlqcdop3}, where
 the lattice data from Refs.~\cite{Primer:2013pva, Parreno:2016fwu} are also shown. One can see
that only the EOMS formulation can describe the lattice QCD data reasonably well, consistent
with the finding in the SU(2) sector~\cite{Pascalutsa:2004ga}. Furthermore,  as shown
in Table~\ref{tab:op3fit}, the LECs of the EOMS formulation determined from the fit to the
lattice QCD data are similar to those determined by the experimental data~\cite{Geng:2008mf}.

 \begin{figure}[htpb]
  \centering
  \includegraphics[width=0.8\textwidth]{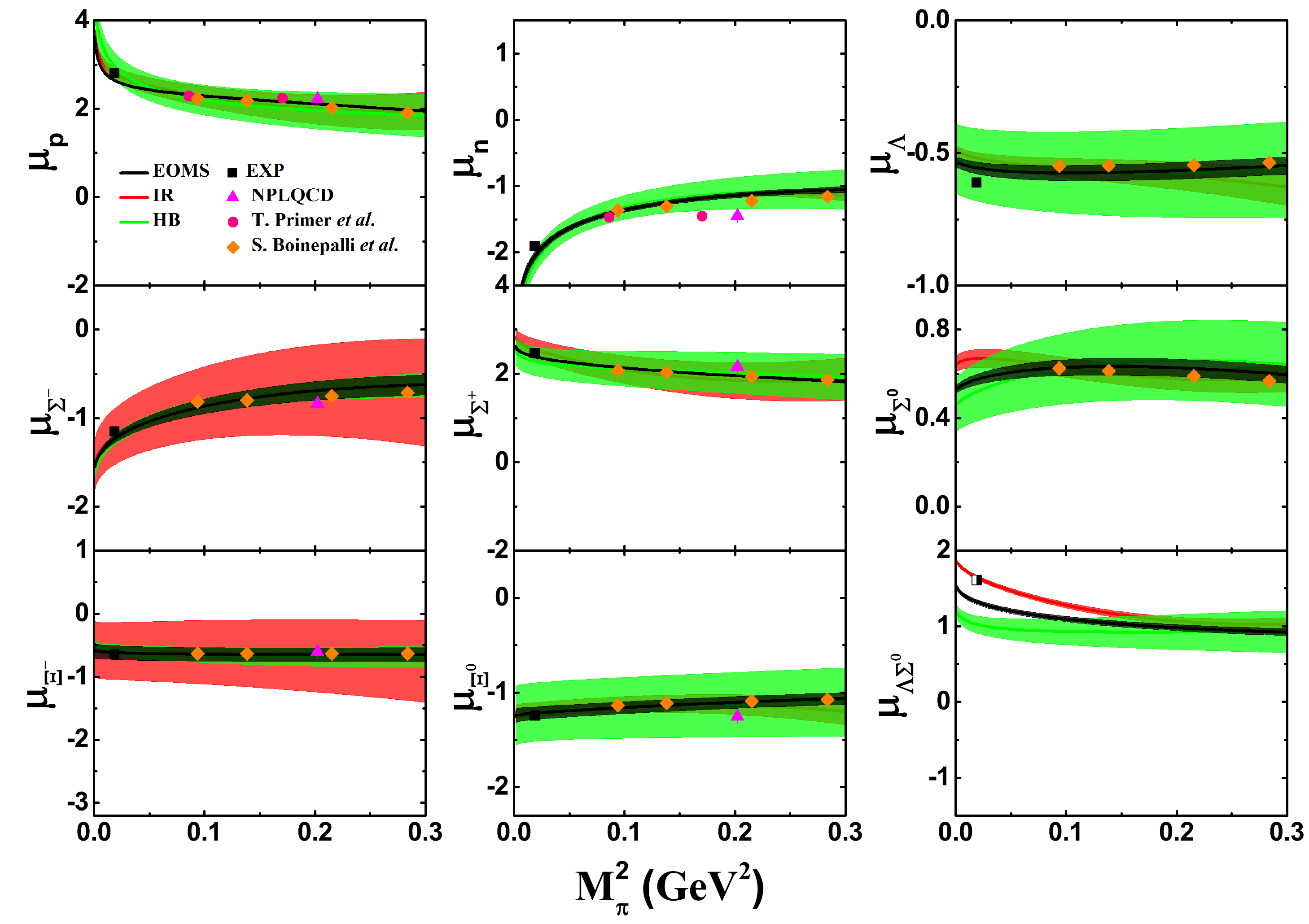}
  \caption{Fits to the lattice QCD data of Ref.~\cite{ Boinepalli:2006xd} at $\mathcal{O}(p^4)$ with the EOMS, HB, and IR BChPT. The bands represent
  higher order contributions as explained in the text.}
  \label{fig:fitlqcdop4}
\end{figure}

\begin{table}[h]
\renewcommand{\tabcolsep}{0.2cm}
\caption{Low-energy constants  obtained by fitting to the lattice QCD data of Ref.~\cite{Boinepalli:2006xd} and the corresponding $\tilde{\chi}^2$ in
the IR, HB, and EOMS BChPT up to NNLO.}
\begin{tabular}{ccccccccccc}
  \hline
  \hline
   &$b_6^D$   &$b_6^F$   &$\alpha_1$    &$\alpha_2$   &$\alpha_3$   &$\alpha_4$   &$\beta_1$   &$b_6^{D'}$    &$b_6^{F'}$   &$\tilde{\chi}^2$\\
  \hline
   IR                &4.02     &2.08     &$-$0.20    &$-$0.83    &0.06    &0.20     &$-$2.88    &$-$3.66    &$-$3.59    &0.14\\
   HB                &2.16     &1.08     &$-$1.47    &0.28     &$-$1.47   &1.53     &$-$2.12    &0.56     &0.89     &0.24\\
   EOMS              &3.03     &1.40     &0.17     &0.30     &0.15    &0.60     &$-$0.56    &$-$0.69    &$-$0.59    &0.13\\
  \hline
  \hline
\end{tabular}\label{tab:op4fit}
\end{table}
The situation becomes different, however, if we fit to the lattice data with the $\mathcal{O}(p^4)$
BChPT results, all the three formulations can
describe the lattice data with similar quality, as shown in Fig.~\ref{fig:fitlqcdop4} and
Table~\ref{tab:op4fit}. On the other hand, it seems that with the LECs determined from the
best fit, the predicted $\mu_{\Lambda\Sigma_0}$ by the IR formulation
is in better agreement with the experimental value (if a positive sign is taken). Nevertheless,
one should note that the convergence pattern, particularly those of the IR and HB schemes,
is  destroyed (see Table~\ref{tab:convlat}). 

\begin{figure}[htpb]
  \centering
  \includegraphics[width=0.8\textwidth]{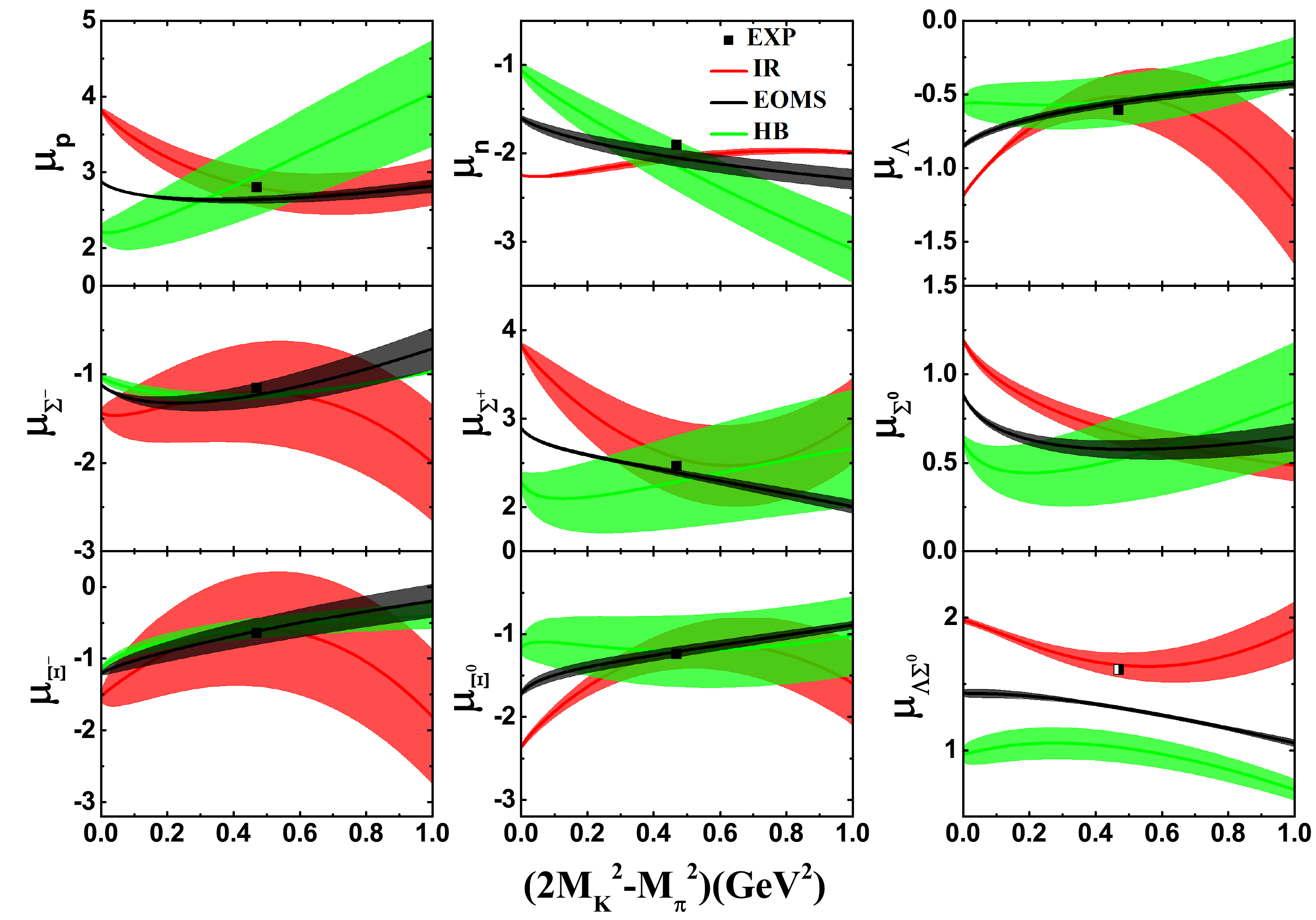}
  \caption{Predicted dependence of the octet baryon magnetic moments on the strange quark mass with the
  LECs determined by fitting to the lattice QCD data of Ref.~\cite{Boinepalli:2006xd} at $\mathcal{O}(p^4)$ with the EOMS, HB, and IR BChPT. The $u/d$ quark mass is
  fixed at their physical value while the strange quark mass is proportional to $2M_K^2-M_\pi^2$ according to leading order ChPT. The bands represent higher order contributions.}
  \label{fig:ms1}
\end{figure}

\begin{figure}[htpb]
  \centering
  \includegraphics[width=0.8\textwidth]{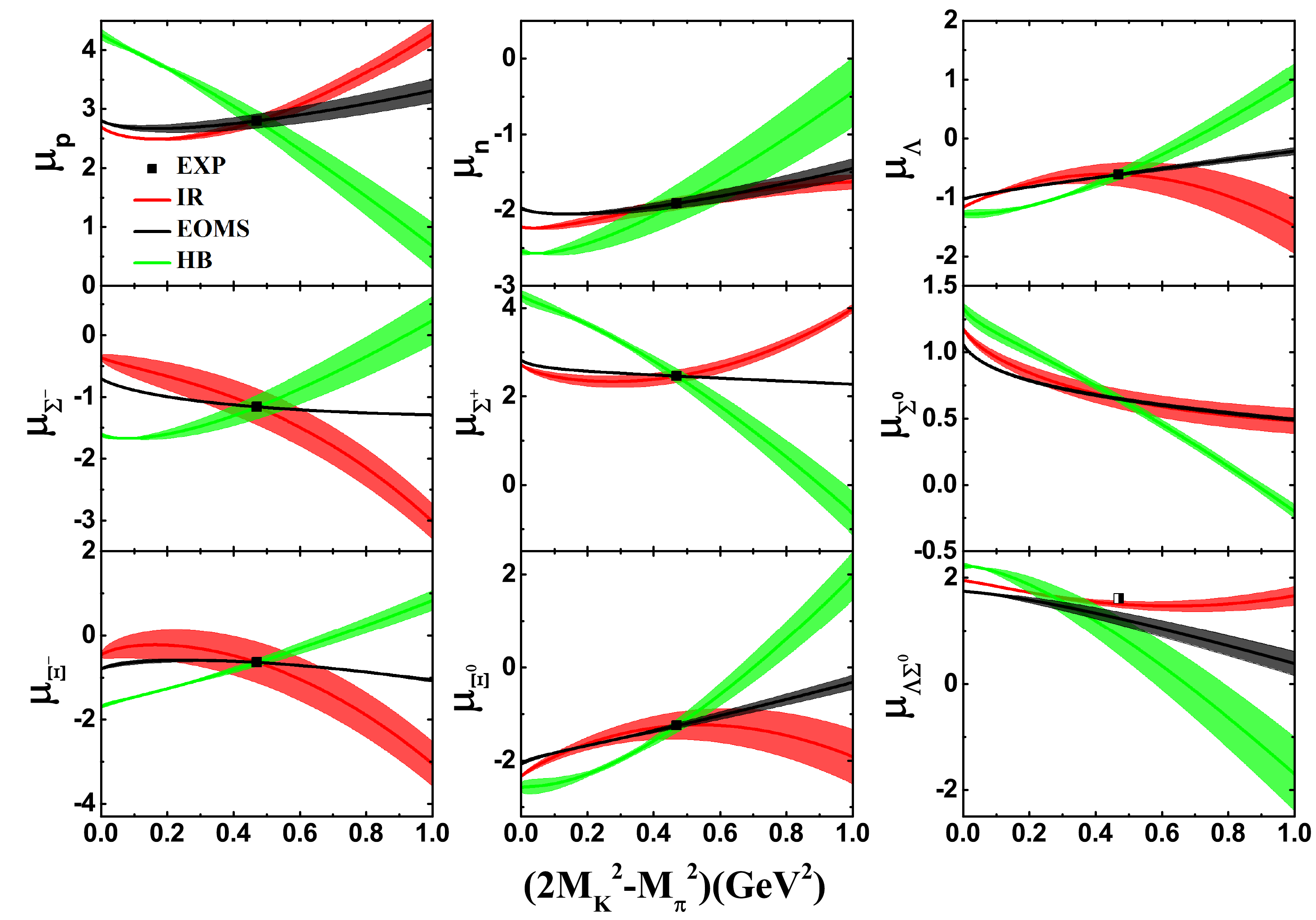}
  \caption{The same as Fig.~6, but with the LECs  given in Table II.}
  \label{fig:ms2}
\end{figure}

 One way to distinguish the different formulations in
the present case is to study the strange quark mass dependence of the magnetic moments,
shown in Figs.~\ref{fig:ms1}, ~\ref{fig:ms2}.  One can see that depending on how one determines
the LECs, either
by fitting to the lattice QCD data or to the experimental data with the convergence constraint,
the dependences on the strange quark mass are quite different.~\footnote{We note that the EOMS predictions in the two cases are similar to each other.} It is clear that one needs  more lattice data with varying strange quark mass to check which scenario is more realistic.

\begin{table}[htpb]
\centering
\renewcommand{\tabcolsep}{0.2cm}
\renewcommand\arraystretch{1.2}
\caption{Contributions of different chiral orders of the HB, IR, and EOMS schemes up to $\mathcal{O}(p^4)$ with the LECs obtained by fitting to the lattice QCD data.}
\begin{tabular}{ccccccc}
 \hline
 \hline
 \multirow{2}{*}{Baryons} &\multicolumn{2}{c}{EOMS}            &\multicolumn{2}{c}{IR}          &\multicolumn{2}{c}{HB}\\
 \cline{2-7}
                          &$\mu_B^{(3)}/\mu_B^{(2)}$        &$\mu_B^{(4)}/\mu_B^{(3)}$             &$\mu_B^{(3)}/\mu_B^{(2)}$         &$\mu_B^{(4)}/\mu_B^{(3)}$           &$\mu_B^{(3)}/\mu_B^{(2)}$          &$\mu_B^{(4)}/\mu_B^{(3)}$  \\
\hline
 $p$                        &$-$0.26          &$-$0.12               &$-$0.12         &2.09            &$-$0.73	        &$-$1.08\\
 $n$                        &$-$0.22	      &$-$1.09	           &$-$0.17	      &0.38            &$-$0.35	        &$-2.36$\\
 $\Lambda$                &$-$0.61	      &$-$0.26	           &$-$0.72	      &$-$0.15	  &$-$1.62	        &$-$0.85\\
 $\Sigma^{-}$             &0.13	          &$-$1.85               &1.02	      &$-$1.30	       &$-$0.41	        &$-$0.81\\
 $\Sigma^{+}$             &$-$0.31          &$-$0.01	           &$-$0.04     &10.21           &$-$1.03	        &$-$0.82\\
 $\Sigma^{0}$             &$-$0.61	      &$-$0.29               &$-$0.72	      &$-$0.31	       &$-$1.62	        &$-$0.82\\
 $\Xi^{-}$                &0.02   	      &$-$28.25	           &1.22	      &$-$1.53	       &$-$1.03	        &$-$0.44\\
 $\Xi^{0}$                &$-$0.64	      &$-$0.38               &$-$0.76	      &$-$0.27	       &$-$1.71	        &$-$0.91\\
 $\Lambda \Sigma^{0}$     &$-$0.37	      &$-$0.33	           &$-$0.38    &$-$0.22	&$-$0.84	        &$-$0.78\\
\hline
\hline
\end{tabular}\label{tab:convlat}
\end{table}

\section{Summary}
\label{sec:summ}
We studied the octet baryon magnetic moments in  baryon chiral perturbation theory with the
extended-on-mass-shell renormalization scheme up to next-to-next-to-leading order.
We determined the low-energy constants following two different strategies, either by fitting to the experimental data with convergence as a further constraint or by fitting to lattice QCD data directly. It was shown that
in the first case the extended-on-mass shell formulation seems to describe better the lattice QCD data, while in the second case, although all three formulations of baryon chiral perturbation theory can
describe the lattice QCD data, they predict rather different strange quark mass dependence.  Clearly more lattice QCD simulations are needed to better understand the situation and the convergence pattern of baryon chiral perturbation theory.

\section*{Acknowledgements}
This work is supported in part by the National Natural Science Foundation of China under Grants No. 11522539, No. 11735003, the China Postdoctoral Science Foundation (2016M600845, 2017T100008), and by DFG and NSFC through funds provided to the
Sino-German  CRC  110  ``Symmetry  and  the  Emergence of Structure in QCD" (NSFC Grant No.  11621131001,
DFG  Grant  No.   TRR110).
The work of UGM was also supported by the CAS President's International Fellowship Initiative (PIFI) (Grant No.  2018DM0034).

\section*{Appendix}
\subsection{NNLO tree level contributions}
Here, we list the tree level contributions at NNLO.
\begin{equation}
\begin{split}
&\mu_{p}^{(4,d)}=\frac{2}{3} \left(3 b_6^{F'} \left(2 M_K^2+M_\pi ^2\right)+b_6^{D'} \left(2 M_K^2+M_\pi ^2\right)\right.\\
&\left.+2 \left(3 \alpha _2 M_K^2+\alpha _3 M_K^2+\alpha _4
   M_K^2-\beta _1 M_K^2+3 \alpha _1 \left(M_K^2-M_\pi ^2\right)-\alpha _3 M_\pi ^2+\beta _1 M_\pi ^2\right)\right),\\
&\mu_{n}^{(4,d)}=-\frac{4}{3} \left(b_6^{D'} \left(2 M_K^2+M_\pi ^2\right)+2 \alpha _4 M_K^2+\beta _1 M_K^2+2 \alpha _3 \left(M_K^2-M_\pi ^2\right)-\beta _1 M_\pi
   ^2\right),\\
&\mu_{\Lambda}^{(4,d)}=-\frac{2}{9} \left(3 b_6^{D'} \left(2 M_K^2+M_\pi ^2\right)+2 \left(\alpha _4 \left(8 M_K^2-5 M_\pi ^2\right)+3 \beta _1 \left(M_K^2-M_\pi
   ^2\right)\right)\right),\\
&\mu_{\Sigma^+}^{(4,d)}=\frac{2}{3} \left(3 b_6^{F'} \left(2 M_K^2+M_\pi ^2\right)+b_6^{D'} \left(2 M_K^2+M_\pi ^2\right)+2 \left(\beta _1 \left(M_\pi ^2-M_K^2\right)+3 \alpha
   _2 M_\pi ^2+\alpha _4 M_\pi ^2\right)\right),\\
&\mu_{\Sigma^-}^{(4,d)}=\frac{2}{3} \left(-3 b_6^{F'} \left(2 M_K^2+M_\pi ^2\right)+b_6^{D'} \left(2 M_K^2+M_\pi ^2\right)-2 \beta _1 M_K^2-6 \alpha _2 M_\pi ^2+2 \alpha _4
   M_\pi ^2+2 \beta _1 M_\pi ^2\right),\\
&\mu_{\Sigma^0}^{(4,d)}=\frac{2}{3} \left(b_6^{D'} \left(2 M_K^2+M_\pi ^2\right)+2 \left(\beta _1 \left(M_\pi ^2-M_K^2\right)+\alpha _4 M_\pi ^2\right)\right),\\
&\mu_{\Lambda \Sigma^0}^{(4,d)}=\frac{2}{\sqrt{3}}\left(b_6^{D'} \left(2 M_K^2+M_\pi ^2\right)+2 \alpha _4 M_\pi ^2\right),\\
&\mu_{\Xi^-}^{(4,d)}=-\frac{2}{3} \left(3 b_6^{F'} \left(2 M_K^2+M_\pi ^2\right)-b_6^{D'} \left(2 M_K^2+M_\pi ^2\right)\right.\\
&\left.-2 \left(-3 \alpha _2 M_K^2-\alpha _3 M_K^2+\alpha _4
   M_K^2-\beta _1 M_K^2+3 \alpha _1 \left(M_K^2-M_\pi ^2\right)+\alpha _3 M_\pi ^2+\beta _1 M_\pi ^2\right)\right),\\
&\mu_{\Xi^0}^{(4,d)}=-\frac{4}{3} \left(b_6^{D'} \left(2 M_K^2+M_\pi ^2\right)+2 \alpha _4 M_K^2+\beta _1 M_K^2-2 \alpha _3 \left(M_K^2-M_\pi ^2\right)-\beta _1 M_\pi^2\right).
\end{split}
\end{equation}
\subsection{Coefficients appearing in the NNLO loop contributions}
Here, we list the coefficients appearing in the loop contributions at NNLO.
\begin{equation}
\begin{split}
&C_{p\pi}^{(4,e)}=-\frac{1}{2}(b_6^F+b_6^D),C_{pK}^{(4,e)}=-b_6^F,C_{n\pi}^{(4,e)}=\frac{1}{2}(b_6^D+b_6^F),
C_{nK}^{(4,e)}=\frac{1}{2}(b_6^D-b_6^F),\\
&C_{\Lambda \pi}^{(4,e)}=0,C_{\Lambda K}^{(4,e)}=\frac{1}{2}b_6^D,C_{\Sigma^+ \pi}^{(4,e)}=-b_6^F,C_{\Sigma^+ K}^{(4,e)}=-\frac{1}{2}(b_6^D+b_6^F),\\
&C_{\Sigma^- \pi}^{(4,e)}=b_6^F,C_{\Sigma^- K}^{(4,e)}=\frac{1}{2}(b_6^F-b_6^D),C_{\Sigma^0 \pi}^{(4,e)}=0,C_{\Sigma^0 K}^{(4,e)}=-\frac{1}{2}b_6^D,\\
&C_{\Xi^- \pi}^{(4,e)}=\frac{1}{2}(b_6^F-b_6^D),C_{\Xi^- K}^{(4,e)}=b_6^F,C_{\Xi^0 \pi}^{(4,e)}=\frac{1}{2}(b_6^D-b_6^F),C_{\Xi^0 K}^{(4,e)}=\frac{1}{2}(b_6^D+b_6^F),\\
&C_{\Lambda \Sigma^0 \pi}^{(4,e)}=-\frac{1}{\sqrt{3}}b_6^D,C_{\Lambda \Sigma K}^{(4,e)}=-\frac{1}{2\sqrt{3}}b_6^D.
\end{split}
\end{equation}
and
\begin{equation}
\begin{split}
&C_{p\pi}^{(4,f)}=-\frac{1}{4}(D+F)^2(b_6^D-b_6^F),C_{pK}^{(4,f)}=\frac{1}{2}\left(\left(\frac{D^2}{9}-2DF+F^2\right)b_6^D
+(D-F)^2 b_6^F\right),\\
&C_{p\eta}^{(4,f)}=\frac{1}{36}(D-3F)^2(b_6^D+3b_6^F),C_{n\pi}^{(4,f)}=\frac{1}{2}(D+F)^2b_6^F,\\
&C_{nK}^{(4,f)}=-\frac{1}{2}\left(\left(-\frac{7}{9}D^2+\frac{2}{3}DF+F^2\right)b_6^D+\left(D-F\right)^2b_6^F\right),
C_{n\eta}^{(4,f)}=-\frac{1}{18}(D-3F)^2b_6^D,\\
&C_{\Lambda\pi}^{(4,f)}=\frac{1}{3}D^2b_6^D,C_{\Lambda K}^{(4,f)}=-\frac{1}{18}(D^2+9F^2)b_6^D+DFb_6^F, C_{\Lambda \eta}^{(4,f)}=-\frac{1}{9}D^2b_6^D\\
&C_{\Sigma^+ \pi}^{(4,f)}=-\frac{1}{9}(D^2+6DF-6F^2)b_6^D+F^2 b_6^F,\\
&C_{\Sigma^+ K}^{(4,f)}=-\frac{1}{6}(D^2+6DF+F^2)b_6^D+\frac{1}{2}(D-F)^2b_6^F,C_{\Sigma^+ \eta}^{(4,f)}=\frac{1}{9}D^2(b_6^D+3b_6^F),\\
&C_{\Sigma^- \pi}^{(4,f)}=-\frac{1}{9}(D^2-6DF-6F^2)b_6^D-F^2b_6^F,\\
&C_{\Sigma^- K}^{(4,f)}=-\frac{1}{6}(D^2-6DF+F^2)b_6^D-\frac{1}{2}(D+F)^2b_6^F,C_{\Sigma^- \eta}^{(4,f)}=\frac{1}{9}D^2(b_6^D-3b_6^F),\\
&C_{\Sigma^0 \pi}^{(4,f)}=-\frac{1}{9}(D^2-6F^2)b_6^D,C_{\Sigma^0 K}^{(4,f)}=-\frac{1}{6}(D^2+F^2)b_6^D-DFb_6^F,\\
&C_{\Sigma^0 \eta}^{(4,f)}=\frac{1}{9}D^2b_6^D,C_{\Xi^- \pi}^{(4,f)}=-\frac{1}{4}(D-F)^2(b_6^D+b_6^F),\\
&C_{\Xi^- K}^{(4,f)}=\frac{1}{2}\left(\frac{1}{9}D^2+2DF+F^2\right)b_6^D-\frac{1}{2}(D+F)^2b_6^F,\\
&C_{\Xi^- \eta}^{(4,f)}=\frac{1}{36}(D+3F)^2(b_6^D-3b_6^F),C_{\Xi^0 \pi}^{(4,f)}=-\frac{1}{2}(D-F)^2b_6^F,\\
&C_{\Xi^0 K}^{(4,f)}=-\frac{1}{2}\left(-\frac{7}{9}D^2-\frac{2}{3}DF+F^2\right)b_6^D+\frac{1}{2}(D+F)^2b_6^F,C_{\Xi^0 \eta}^{(4,f)}=-\frac{1}{18}(D+3F)^2b_6^D,\\
&C_{\Lambda \Sigma^0 \pi}^{(4,f)}=\frac{1}{3\sqrt{3}}(D^2b_6^D-6DFb_6^F),C_{\Lambda \Sigma^0 K}^{(4,f)}=\frac{1}{2\sqrt{3}}(3F^2-D^2)b_6^D-\frac{1}{\sqrt{3}}DFb_6^F,\\
&C_{\Lambda \Sigma^0 \eta}^{(4,f)}=-\frac{1}{3\sqrt{3}}D^2b_6^D.
\end{split}
\end{equation}
The wave function renormalization coefficients are invariant under SU(2) transformations, therefore we give only the values of the different multiplets
\begin{equation}
\begin{split}
&C_{N \pi}^{(4,g)}=\frac{3}{4}(D+F)^2,C_{N K}^{(4,g)}=\frac{5}{6}D^2-DF+\frac{3}{2}F^2,C_{N \eta}^{(4,g)}=\frac{1}{12}(D-3F)^2,\\
&C_{\Sigma \pi}^{(4,g)}=\frac{1}{3}D^2+2F^2,C_{\Sigma K}^{(4,g)}=D^2+F^2,C_{\Sigma \eta}^{(4,g)}=\frac{1}{3}D^2,\\
&C_{\Lambda \pi}^{(4,g)}=D^2,C_{\Lambda K}^{(4,g)}=\frac{1}{3}D^2+3F^2,C_{\Lambda \eta}^{(4,g)}=\frac{1}{3}D^2,\\
&C_{\Xi \pi}^{(4,g)}=\frac{3}{4}(D-F)^2,C_{\Xi K}^{(4,g)}=\frac{5}{6}D^2+DF+\frac{3}{2}F^2,C_{\Xi \eta}^{(4,g)}=\frac{1}{12}(D+3F)^2.
\end{split}
\end{equation}
and
\begin{equation}
\begin{split}
&C_{p \pi}^{(4,h)}=2(b_{10}+b_{11}),C_{p K}^{(4,h)}=b_{9}+4b_{11},C_{n \pi}^{(4,h)}=-2(b_{10}+b_{11}),C_{n K}^{(4,h)}=-2b_{10}+2b_{11},\\
&C_{\Lambda \pi}^{(4,h)}=0,C_{\Lambda K}^{(4,h)}=-2b_{10},C_{\Sigma^+ \pi}^{(4,h)}=b_{9}+4b_{11},C_{\Sigma^+ K}^{(4,h)}=2(b_{10}+b_{11}),\\
&C_{\Sigma^- \pi}^{(4,h)}=-b_{9}-4b_{11},C_{\Sigma^- K}^{(4,h)}=2(b_{10}-b_{11}),C_{\Sigma^0 \pi}^{(4,h)}=0,C_{\Sigma^0 K}^{(4,h)}=2b_{10},\\
&C_{\Xi^- \pi}^{(4,h)}=2(b_{10}-b_{11}),C_{\Xi^- K}^{(4,h)}=-b_{9}-4b_{11},C_{\Xi^0 \pi}^{(4,h)}=-2b_{10}+2b_{11},\\
&C_{\Xi^0 K}^{(4,h)}=-2(b_{10}+b_{11}),C_{\Lambda \Sigma^0 \pi}^{(4,h)}=\frac{4}{\sqrt{3}}b_{10},C_{\Lambda \Sigma^0 K}^{(4,h)}=\frac{2}{\sqrt{3}}b_{10}.\\
\end{split}
\end{equation}
and
\begin{equation}
\begin{split}
&C_{p \pi}^{(4,i)}=-({D}+{F})^2 \left({b_D}
   {M_K}^2+{b_F} \left({M_\pi}^2-{M_K}^2\right)\right),C_{p K}^{(4,i)}=-2 {b_D} {M_\pi }^2 ({D}-{F})^2,\\
&C_{p \eta}^{(4,i)}=-\frac{1}{3} ({D}-3 {F})^2 \left({b_D}{M_K}^2+{b_F} \left({M_\pi}^2-{M_K}^2\right)\right),\\
&C_{n \pi}^{(4,i)}=-2 ({D}+{F})^2 \left({M_K}^2
   ({b_D}-{b_F})+{b_F} {M_\pi}^2\right),C_{n K}^{(4,i)}=2 {b_D} {M_\pi }^2 ({D}-{F})^2,\\
&C_{n \eta}^{(4,i)}=0,\\
&C_{\Lambda \pi}^{(4,i)}=0,C_{\Lambda K}^{(4,i)}=-\frac{2}{3} \left(6 {b_D} {D} {F}
   {M_K}^2-{b_F} \left({D}^2+9{F}^2\right) ({M_K}-{M_\pi })({M_K}+{M_\pi })\right),C_{\Lambda \eta}^{(4,i)}=0,\\
&C_{\Sigma^+ \pi}^{(4,i)}=-4 {b_D} {F}^2 {M_\pi}^2,C_{\Sigma^+K}^{(4,i)}=-2 ({D}-{F})^2 \left({M_K}^2
   ({b_D}-{b_F})+{b_F} {M_\pi}^2\right),C_{\Sigma^+\eta}^{(4,i)}=-\frac{4}{3} {b_D} {D}^2 {M_\pi }^2\\
&C_{\Sigma^- \pi}^{(4,i)}=4 {b_D} {F}^2 {M_\pi }^2,C_{\Sigma^- K}^{(4,i)}=2 ({D}+{F})^2 \left({M_K}^2({b_D}+{b_F})-{b_F} {M_\pi}^2\right),C_{\Sigma^- \eta}^{(4,i)}=\frac{4}{3} {b_D} {D}^2 {M_\pi }^2\\
&C_{\Sigma^0 \pi}^{(4,i)}=0,C_{\Sigma^0 K}^{(4,i)}=2 \left(2 {b_D} {D} {F}{M_K}^2+{b_F}
   \left({D}^2+{F}^2\right)({M_K}-{M_\pi }) ({M_K}+{M_\pi})\right),\\ &C_{\Sigma^0 K}^{(4,i)}=0,\\
&C_{\Xi^- \pi}^{(4,i)}=({D}-{F})^2 \left({b_D}
   {M_K}^2+{b_F} \left({M_K}^2-{M_\pi}^2\right)\right),\\
&C_{\Xi^- K}^{(4,i)}=2 {b_D} {M_\pi }^2 ({D}+{F})^2,\\
&C_{\Xi^- \eta}^{(4,i)}=\frac{1}{3} ({D}+3 {F})^2 \left({b_D}
   {M_K}^2+{b_F} \left({M_K}^2-{M_\pi}^2\right)\right),\\
&C_{\Xi^0 \pi}^{(4,i)}=2 ({D}-{F})^2 \left({M_K}^2
   ({b_D}+{b_F})-{b_F} {M_\pi}^2\right),C_{\Xi^0 K}^{(4,i)}=-2 {b_D} {M_\pi }^2 ({D}+{F})^2,\\
&C_{\Xi^0 \eta}^{(4,i)}=0,\\
&C_{\Lambda\Sigma^0 \pi}^{(4,i)}={8/{\sqrt{3}} {b_D} {D} {F} {M_\pi
   }^2},C_{\Lambda\Sigma^0 K}^{(4,i)}={2/{\sqrt{3}} \left(2 {b_D} {D} {F}
   {M_K}^2-{b_F} \left({D}^2-3{F}^2\right) \left({M_K}^2-{M_\pi}^2\right)\right)},\\
&C_{\Lambda\Sigma^0 \eta}^{(4,i)}=0.
\end{split}
\end{equation}
and
\begin{equation}
\begin{split}
&C_{p \pi}^{(4,j)}=2(D+F)^2\left(M_K^2 b_D+\left(M_\pi^2-M_{K}^2\right)b_F\right),\\
&C_{p K}^{(4,j)}=\frac{1}{3}\left(\left(3F+D\right)^2M_{\eta}^2+3(D-F)^2M_{\pi}^2\right)b_D,\\
&C_{n \pi}^{(4,j)}=-2(D+F)^2\left(M_{K}^2b_D+\left(M_{\pi}^2-M_{K}^2\right)b_F\right),C_{n K}^{(4,j)}=2(D-F)^2M_{\pi}^2b_D,\\
&C_{\Lambda \pi}^{(4,j)}=0,C_{\Lambda K}^{(4,j)}=-4DFM_{K}^2b_D+\frac{2}{3}(9F^2+D^2)(M_{K}^2-M_{\pi}^2)b_F,\\
&C_{\Sigma^+ \pi}^{(4,j)}=\frac{4}{3}(3F^2M_{\pi}^2+D^2M_{\eta}^2)b_D,C_{\Sigma^+ K}^{(4,j)}=2(D+F)^2\left(M_{K}^2b_D-\left(M_{\pi}^2-M_{K}^2\right)b_F\right),\\
&C_{\Sigma^- \pi}^{(4,j)}=-\frac{4}{3}(3F^2M_{\pi}^2+D^2M_{\eta}^2)b_D,\\
&C_{\Sigma^- K}^{(4,j)}=-2(D-F)^2\left(M_{K}^2b_D+\left(M_{\pi}^2-M_{K}^2\right)b_F\right),\\
&C_{\Sigma^0 \pi}^{(4,j)}=0,C_{\Sigma^0 K}^{(4,j)}=4DFM_{K}^2b_D+2(D^2+F^2)\left(M_{K}^2-M_{\pi}^2\right)b_F,\\
&C_{\Xi^- \pi}^{(4,j)}=-2(D-F)^2\left(M_{K}^2b_D-\left(M_{\pi}^2-M_{K}^2\right)b_F\right),\\
&C_{\Xi^- K}^{(4,j)}=-\frac{4}{9} {b_D} \left({M_K}^2 ({D}-3
   {F})^2+2 {D} {M_\pi }^2
   ({D}+3 {F})\right),\\
&C_{\Xi^0 \pi}^{(4,j)}=2(D-F)^2\left(M_{K}^2b_D+\left(M_{K}^2-M_{\pi}^2\right)b_F\right),C_{\Xi^0 K}^{(4,j)}=-2(D-F)^2M_{\pi}^2b_D,\\
&C_{\Lambda\Sigma^0 \pi}^{(4,j)}=8/\sqrt{3}DFM_{\pi}^2b_D,C_{\Lambda\Sigma^0 \pi}^{(4,j)}=4/\sqrt{3}DFM_{K}^2b_D-2/\sqrt{3}(3F^2-D^2)(M_{\pi}^2-M_{K}^2)b_F.
\end{split}
\end{equation}


\begin{thebibliography}{99}




\bibitem{Coleman:1961jn}
  S.~R.~Coleman and S.~L.~Glashow,
  Phys.\ Rev.\ Lett.\  {\bf 6}, 423 (1961).

\bibitem{Gasser:1984gg}
  J.~Gasser and H.~Leutwyler,
  Nucl.\ Phys.\  B {\bf 250}, 465 (1985).

\bibitem{Gasser:1987rb}
  J.~Gasser, M.~E.~Sainio and A.~Svarc,
  Nucl.\ Phys.\  B {\bf 307}, 779 (1988).

\bibitem{Bernard:1995dp}
  V.~Bernard, N.~Kaiser and U.-G.~Mei{\ss}ner,
  Int.\ J.\ Mod.\ Phys.\ E {\bf 4}, 193 (1995)

\bibitem{Scherer:2002tk}
  S.~Scherer,
  Adv.\ Nucl.\ Phys.\  {\bf 27}, 277 (2003).

\bibitem{Bernard:2007zu}
  V.~Bernard,
  Prog.\ Part.\ Nucl.\ Phys.\  {\bf 60}, 82 (2008)

\bibitem{Caldi:1974ta}
  D.~G.~Caldi and H.~Pagels,
  Phys.\ Rev.\  D {\bf 10}, 3739 (1974).

\bibitem{Bijnens:1985kj}
  J.~Bijnens, H.~Sonoda and M.~B.~Wise,
  Nucl.\ Phys.\ B {\bf 261}, 185 (1985).

\bibitem{Jenkins:1992pi}
  E.~E.~Jenkins, M.~E.~Luke, A.~V.~Manohar and M.~J.~Savage,
  Phys.\ Lett.\  B {\bf 302}, 482 (1993), [Erratum-ibid.\  B {\bf 388} (1996) 866].

\bibitem{Kubis:2000aa}
  B.~Kubis and U.-G.~Mei{\ss}ner,
  Eur.\ Phys.\ J.\ C {\bf 18}, 747 (2001)

\bibitem{Donoghue:2004vk}
  J.~F.~Donoghue, B.~R.~Holstein, T.~Huber and A.~Ross,
  Fizika B {\bf 14}, 217 (2005).

\bibitem{Geng:2008mf}
  L.~S.~Geng, J.~Martin Camalich, L.~Alvarez-Ruso and M.~J.~Vicente Vacas,
  Phys.\ Rev.\ Lett.\  {\bf 101}, 222002 (2008)


  \bibitem{Fuchs:2003qc}
   J.~Gegelia and G.~Japaridze,
  Phys.\ Rev.\  D {\bf 60}, 114038 (1999); T.~Fuchs, J.~Gegelia, G.~Japaridze and S.~Scherer,
  Phys.\ Rev.\  D {\bf 68} (2003) 056005.

\bibitem{Geng:2009hh}
  L.~S.~Geng, J.~Martin Camalich and M.~J.~Vicente Vacas,
  Phys.\ Lett.\ B {\bf 676}, 63 (2009).
  
 
\bibitem{Bernard:1998gv} 
  V.~Bernard, H.~W.~Fearing, T.~R.~Hemmert and U.-G.~Mei\ss ner,
  Nucl.\ Phys.\ A {\bf 635}, 121 (1998)
  Erratum: [Nucl.\ Phys.\ A {\bf 642}, 563 (1998)].

 

\bibitem{Durand:1997ya}
  L.~Durand and P.~Ha,
  Phys.\ Rev.\  D {\bf 58} (1998) 013010.

\bibitem{Puglia:1999th}
  S.~J.~Puglia and M.~J.~Ramsey-Musolf,
  Phys.\ Rev.\  D {\bf 62}, 034010 (2000).

\bibitem{Meissner:1997hn}
  U.-G.~Mei{\ss}ner and S.~Steininger,
  Nucl.\ Phys.\  B {\bf 499}, 349 (1997).




\bibitem{Jenkins:1990jv}
  E.~E.~Jenkins and A.~V.~Manohar,
  Phys.\ Lett.\ B {\bf 255}, 558 (1991).



\bibitem{Bernard:1992qa}
  V.~Bernard, N.~Kaiser, J.~Kambor and U.-G.~Mei{\ss}ner,
  Nucl.\ Phys.\ B {\bf 388}, 315 (1992).

\bibitem{Becher:1999he}
  T.~Becher and H.~Leutwyler,
  Eur.\ Phys.\ J.\  C {\bf 9} (1999) 643.

\bibitem{Geng:2013xn}
  L.~S.~Geng,
  Front.\ Phys.\ (Beijing) {\bf 8}, 328 (2013)

\bibitem{Jiang:2016vax}
  S.~Z.~Jiang, Q.~S.~Chen and Y.~R.~Liu,
  Phys.\ Rev.\ D {\bf 95}, 014012 (2017)

\bibitem{Binder:2015mbz}
  S.~Binder {\it et al.} [LENPIC Collaboration],
  Phys.\ Rev.\ C {\bf 93}, 044002 (2016).
  
%
%
%

\bibitem{Boinepalli:2006xd}
  S.~Boinepalli, D.~B.~Leinweber, A.~G.~Williams, J.~M.~Zanotti and J.~B.~Zhang,
  Phys.\ Rev.\ D {\bf 74}, 093005 (2006)

\bibitem{Primer:2013pva}
  T.~Primer, W.~Kamleh, D.~Leinweber and M.~Burkardt,
  Phys.\ Rev.\ D {\bf 89}, 034508 (2014)

\bibitem{Parreno:2016fwu}
  A.~Parreno, M.~J.~Savage, B.~C.~Tiburzi, J.~Wilhelm, E.~Chang, W.~Detmold and K.~Orginos,
  Phys.\ Rev.\ D {\bf 95},  114513 (2017)


\bibitem{Pascalutsa:2004ga}
  V.~Pascalutsa, B.~R.~Holstein and M.~Vanderhaeghen,
  Phys.\ Lett.\ B {\bf 600}, 239 (2004)






%




\end{thebibliography}
\end{document}